\documentclass[final,3p,times]{elsarticle}
\usepackage{}

\usepackage{amssymb}
\usepackage{amsmath}
\usepackage{lscape}

\usepackage{varwidth}
\usepackage{multirow}
\usepackage{dcolumn}
\usepackage{tabularx}
\usepackage{booktabs}
\newcolumntype{C}{>{\centering\arraybackslash}X}

\usepackage{graphicx}
\usepackage{epstopdf}
\usepackage{graphics}
\usepackage{graphicx}
\usepackage{epsfig}
\usepackage{epstopdf}

\newcounter{bla}

\journal{Computer Physics Communications}

\newcommand{\Ncoll}{\ensuremath{N_{\rm coll}}}
\newcommand{\Npart}{\ensuremath{N_{\rm part}}}
\newcommand{\sNN}{\ensuremath{\sqrt{s_{\rm NN}}}}
\newcommand{\pT}{\ensuremath{p\rm {_T}}}

\begin{document}

\begin{frontmatter}



\title{Revisiting the centrality definition and observable centrality
dependence \\of relativistic heavy-ion collisions in PACIAE model}


\author[a]{Yu-Liang Yan\corref{author}}
\author[b]{Dai-Mei Zhou}
\author[b]{An-Ke Lei}
\author[a]{Xiao-Mei Li}
\author[b]{Xiao-Ming Zhang}
\author[c]{Liang Zheng}
\author[c]{Gang Chen}
\author[b]{Xu Cai}
\author[a,b]{Ben-Hao Sa}

\cortext[author] {Corresponding author.\\\textit{E-mail address:} yanyl@ciae.ac.cn}
\address[a]{China Institute of Atomic Energy, P. O. Box 275 (10),
              Beijing, 102413 China.}
\address[b]{Institute of Particle Physics, Central China Normal
              University, 430082 Wuhan, China \\ and Key Laboratory of Quark
              and Lepton Physics (CCNU), Ministry of Education, China.}
\address[c]{School of Mathematics and Physics, China University of Geosciences (Wuhan),
	\break Wuhan 430074, China}

\begin{abstract}
We improve the centrality definition in impact parameter in PACIAE model
responding the fact reported by the ALICE, ATLAS, and CMS collaborations
that the maximum impact parameter in heavy ion collisions should be extended to
20 $fm$. Meanwhile the PACIAE program is updated to a new version of
PACIAE 2.2.2 with convenience of studying the elementary nuclear collisions,
proton-nucleus collisions, and the nucleus-nucleus collisions in one unified
program version. The new impact parameter definition together with the
optical Glauber model calculated impact parameter bin, $\Npart$, and $\Ncoll$
in proton-nucleus and nucleus-nucleus collisions at relativistic energies are
consistent with the improved MC-Glauber model ones within the error bar. The
charged-particle pseudorapidity and the transverse momentum distributions in
Pb-Pb collisions at $\sNN=5.02$~TeV simulated by PACIAE 2.2.2 well reproduce
the ALICE experimental data.
\end{abstract}


\end{frontmatter}



{\bf PROGRAM SUMMARY/NEW VERSION PROGRAM SUMMARY}

\begin{small}
\noindent
{\em Program Title: PACIAE version 2.2.2}                  \\
{\em CPC Library link to program files:} (to be added by Technical Editor) \\
{\em Code Ocean capsule:} (to be added by Technical Editor)\\
{\em Licensing provisions: CC By 4.0}                       \\
{\em Programming language: FORTRAN}                       \\
{\em Supplementary material:}                                 \\
{\em Journal reference of previous version: Comput. Phys. Comm.
224 (2018) 417}                  \\
 {\em Does the new version supersede the previous version?: Yes}   \\
 {\em Reasons for the new version: Recently ALICE, ATLAS and CMS collaborations
reported that the maximum impact parameter $b_{max}$ should be extended to 20
$fm$ in the nuclear-nuclear collisions at relativistic energies. The impact
parameter formula in PACIAE model has to be improved correspondingly. Meanwhile
the PACIAE model is updated to PACIAE 2.2.2 with the convenience of studying
the elementary nuclear collisions, proton-nucleus collisions, and the
nucleus-nucleus collisions in one unified program version.}\\
 {\em Summary of revisions:
The impact parameter $b$ in PACIAE model is calculated by geometrical
model of
\begin{equation}
b=\sqrt{c}\times b_{max}, \nonumber
\end{equation}
where $c$ refers to the centrality percentile and $b_{max}$ is assumed to be
\begin{equation}
b_{max}=R_{A}+ R_{B}+f\times d. \nonumber
\end{equation}
In above equation $R_{A}$ ($R_{B}$) is the radius of nuclear $A$ ($B$),
$d=0.546 fm$ describes the tail of the nuclear density profile. Originally,
the coefficient $f$ is set to be equal to 2 and 1 for proton-nucleus and
nucleus-nucleus collisions, respectively. Now they are assumed to be equal to
4 and 2, respectively. Meanwhile, the PACIAE model is updated to the version
of PACIAE 2.2.2 with the convenience of studying the elementary nuclear
collisions, proton-nucleus collisions, and the nucleus-nucleus collisions in
an unified program version.}\\
{\em Nature of problem:
The ALICE,  ATLAS and CMS collaborations reported that the maximum impact
parameter, $b_{max}$, in heavy-ion collisions at relativistic energies should
be extended to 20 $fm$ where the interaction is really approaching to zero.
The impact parameter centrality determination in heavy-ion collisions at
relativistic energies has to be revised accordingly in the PACIAE model.} \\
{\em Solution method:
A new $f$ coefficient set in the impact parameter formula in PACIAE model,
sequentially the new $b$ bin corresponding to a given centrality percentile
bin are introduced in the new version of PACIAE 2.2.2.
} \\
{\em Additional comments including Restrictions and Unusual features:
Depend on the problem studied.}\\

\end{small}

\section{Introduction}
The ultra-relativistic heavy-ion collisions produce strongly interacting
quark-gluon matter (sQGM) under extreme conditions of temperature and energy
density at Relativistic Heavy Ion Collider (RHIC) \cite{rhic1,rhic2,rhic3,rhic4}
and Large Hadron Collider (LHC) \cite{ALICE1,CMS1,ATLAS1}. The centrality is
one key physical character in studying the high energy heavy-ion collisions,
because it is directly related to the interacting volume (overlap zone) of the
collision system. This overlap zone depends on the impact parameter $b$
defined as the distance between the centers of two colliding nuclei in the
plane transverse to the beam axis. The centrality of a nucleus-nucleus (AA)
collision with impact parameter $b$ is usually defined as a percentile $c$ in
the nucleus-nucleus total cross section $\sigma_{AA}$ \cite{abel1}:
\begin{equation}
c=\frac{\int_0^b d\sigma/db^{'}db^{'}}{\int_0^{\infty} d\sigma/db^{'} db^{'}}
 =\frac{1}{\sigma_{AA}}\int_0^b d\sigma/db^{'} db^{'}.
\label{bas1}
\end{equation}

In the experiment, this centrality percentile $c$ of the nucleus-nucleus total
cross section is usually assumed to be approximately equivalent to the
fraction of charged particle multiplicity above a multiplicity cut of
$N_{ch}^{cut}$, or to the energy deposited in the zero-degree calorimeter
(ZDC) below a cut of $E_{ZDC}^{cut}$\cite{abel1}:
\begin{equation}
\begin{aligned}
c\approx &\frac{1}{N_{ch}^{tot}}\int_{N_{ch}^{cut}}^{N_{ch}^{tot}}
d\sigma/dN_{ch}^{'} dN_{ch}^{'} \\
\approx &\frac{1}{E_{ZDC}^{tot}}\int_0^{E_{ZDC}^{cut}}
d\sigma/dE_{ZDC}^{'} dE_{ZDC}^{'}.
\label{bas3}
\end{aligned}
\end{equation}
The nucleus-nucleus total cross section in Eq. (\ref{bas1}) is calculated by
\begin{equation}
\sigma_{AA}=\pi b_{max}^2 \times \frac{N_{evt}(N_{nn-c}\geq 1)}
	{N_{evt}(N_{nn-c}\geq 0)},
\label{bas2}
\end{equation}
i.e. by the nucleus-nucleus geometrical total cross section ($\pi b_{max}^2$)
corrected with the fraction of events with at least one nucleon-nucleon
collision. Meanwhile, the centrality percentile $c$ in a nucleus-nucleus
collision is also assumed to be equivalent to the fraction of impact parameter
distribution ($f(b)\propto bdb$). Therefore, a mapping relation of
\begin{equation}
b=\sqrt c \times b_{max}
\label{eq1}
\end{equation}
is obtained. In above equation $c$ refers to the centrality percentile and
$b_{max}$ is assumed to be
\begin{equation}
b_{max}=R_{A}+ R_{B}+f\times d,
\label{eq2}
\end{equation}
where the $R_{A}$, for instance, is radius of nucleus $A$ (it denotes the
atomic number of nucleus either), the $d=0.546 fm$ refers to the tail of the
nuclear density profile, and the coefficient $f$ is a free parameter.
\begin{equation}
R_A=r_0A^{1/3}, \hspace{0.2cm} r_0=1.12 fm,
\end{equation}

\section{Methodology}
The PACIAE model \cite{sa,sa2,zhou} is a parton and hadron cascade model
based on PYTHIA \cite{PYTHIA}. For nucleon-nucleon (NN) collisions, with
respect to PYTHIA, the partonic and hadronic rescatterings are introduced
before and after the hadronization, respectively. The final
hadronic state is developed from the initial partonic hard scattering and
parton showers, followed by parton rescattering, string fragmentation, and
hadron rescattering stages. Thus, the PACIAE model provides a multi-stage
transport description on the evolution of the collision system.

For a nucleus-nucleus (AA) collisions, the initial positions of nucleons in
the colliding nuclei are sampled according to the Woods-Saxon distribution.
Together with the initial momentum setup of $p_{x}=p_{y}= 0$ and
$p_{z}=p_{\rm beam}$ for each nucleon, a list containing the initial state of
all nucleons in AA collision is constructed. A collision happens between two
nucleons from different nucleus if their relative transverse distance is less
than or equal to the minimum approaching distance:
$D\leq\sqrt{\sigma_{\rm NN}^{\rm tot}/\pi}$. The collision time is calculated
with the assumption of straight-line trajectories. All such nucleon pairs
compose a nucleon-nucleon (NN) collision time list. The NN collision with
least collision time is selected from the list and executed by PYTHIA (PYEVNW
subroutine) with the hadronization temporarily turned-off, as well as the
strings and diquarks broken-up. The nucleon list and NN collision time list
are then updated. A new NN collision with least collision time is selected
from the updated NN collision time list and executed by PYTHIA. With repeating
the aforementioned steps till the NN collision list empty, the initial
partonic state is constructed for a AA collision.

Then, the partonic rescatterings are performed, where the LO-pQCD
parton-parton cross section~\cite{Combridge,Field} is employed. After partonic
rescattering, the string is recovered and then hadronized with the Lund string
fragmentation scheme resulting in an intermediate hadronic state. Finally, the
system proceeds into the hadronic rescattering stage and produces the final
hadronic state observed in the experiments.

Thus PACIAE Monte-Carlo simulation provides a complete description of the NN
and/or AA collisions, which includes the partonic initialization stage,
partonic rescattering stage, hadronization stage, and the hadronic
rescattering stage. Meanwhile, the PACIAE model simulation could be selected
to stop at any stage desired conveniently. In this work, the simulations are
stopped after the hadronic rescattering stage, i.e. at the final hadronic
state.

In the PACIAE 2.0 and successors, we introduce the geometrical model of
Eq.(\ref{eq1}) and set the coefficient of $f$ to be equal to 2 and 1 for
nucleus-nucleus and proton nucleus collisions~\cite{sa}, respectively. The
results of this geometrical model were very consistent with the STAR/RHIC
centrality definitions~\cite{sa,star}, because of they have the same $b_{max}$
definition. Later on, the ALICE, ATLAS and CMS collaborations at LHC have
observed that the $b_{max}$ should be extended to 20 $fm$, where the
interaction is just approaching to zero~\cite{abel1}. This 20 $fm$ is
larger than the $b_{max}$ calculated by Eq.(\ref{eq1}) with above $f$
coefficient setting. A new set of $f$ coefficient is then required.

In the improved Monte Carlo Glauber (MC-Glauber, MCG) model simulation
\cite{loiz} the impact parameter $b$ bin, corresponding to a given centrality
percentile bin, is sliced according to impact parameter distribution up to a
$b_{cut}$, which is assumed to be at somewhere between $R_{A}+R_{B}$ and
20 $fm$. But the last $b$ bin is just assumed to be extended from $b_{cut}$ to
the $20 fm$. For the Pb-Pb, Xe-Xe, Au-Au and Cu-Cu collisions at the
relativistic energies, the $b_{cut}$ is assumed to be approximately equal to
15.6, 13.8, 14.9, and 11.0 $fm$ \cite{loiz}, respectively. Substituting these
$b_{cut}$ values into left side of Eq.(\ref{eq2}) individually, the
corresponding $f$ value of 4.21, 4.45, 3.42, and 3.74 is obtained sequentially.
Therefore we assume the parameter $f$ in Eq. (\ref{eq2}) equals to 4 for the
nucleus-nucleus collision but equals to 2 for the proton-nucleus collision, 
and we just slice the last $b$ bin up to $b_{cut}$ in this work.

The participant nucleons number $\Npart$ and  binary nucleon-nucleon
collisions number $\Ncoll$ are commonly used to represent the collision
centrality. In MC-Glauber model, they are counted as the number of binary
nucleon collisions happened and the number of nucleons which suffer at least
one collision (wounded nucleons), respectively, in a nucleus-nucleus collision
simulation within the boundary of $b_{cut}$. The $\langle N_{\rm{part}}\rangle$
and $\langle N_{\rm{coll}}\rangle$ denote their average value over events. And
the nuclear overlap function $\langle T_{AA}\rangle$ is calculated by
\begin{equation}
\langle T_{AA}\rangle=\frac{\langle N_{\rm{coll}}\rangle}{\sigma_{NN}^{inel}},	
\label{tanc}
\end{equation}	
where $\sigma_{NN}^{inel}$ is the inelastic nucleon-nucleon (NN) cross section.

In the optical Glauber model\cite{esk} used in PACIAE, the $\Npart$ and
$T_{AB}$ are analytically calculated by
\begin{equation}
\begin{aligned}
N_{part}(b)=&\int T_A(\vec b-\vec s)[1-\exp{(-\sigma_{in}T_B(\vec s))}]d^2s \\
&+ \int T_B(\vec s)[1-\exp{(-\sigma_{in}T_A(\vec b-\vec s))}]d^2s,\\ 	
T_{AB}=&\int T_A(\vec b-\vec s)T_B(\vec s)d^2s, \hspace{0.8cm}\\
T_A(\vec s)=&\int \rho(\vec s,z)dz.
\label{bas7}
\end{aligned}
\end{equation}
for the asymmetric AB collisions. In Eq.(\ref{bas7}) the $\vec s$ refers to a
vector of an area perpendicular to the beam axis $z$, $s=|\vec s|$, and
$\rho(\vec s,z)$ stands for the nuclear density in the volume element of
$d^2sdz$ at point of $(\vec s, z)$. The nuclear density distribution in a
nucleus is assumed to be the spherically symmetric Woods-Saxon density
distribution~\cite{star}
\begin{equation}
\rho(r)=\rho_0[1+\rm{exp}(\frac{r-R_A}{d})]^{-1} \hspace{0.2cm}.
\end{equation}
One assumes further:
\begin{equation}
\rho_0=\frac{A}{\frac{4\pi}{3}R_A^3}
\label{bas11}
\end{equation}
\cite{Shalid}. The notarization
relation of
\begin{equation}
4\pi\int \rho(r)r^2dr=A. \nonumber
\end{equation}
is then required.

\section{Results}

The relation of Eq.~\ref{eq2} with new set of $f$ coefficient, together with
the Eqs.~\ref{bas7}-\ref{bas11} are used in the PACIAE 2.2.2 to calculate the
$b_{max}$ and $b$ bins as well as $\Ncoll$ and $\Npart$ in the Pb-Pb and
p-Pb collisions at $\sqrt{s_{NN}}$=5.02 TeV, Xe-Xe collisions at
$\sqrt{s_{NN}}$=5.44 TeV, as well as Au-Au and the Cu-Cu collisions at
$\sqrt{s_{NN}}$=0.2 TeV, respectively. The nucleon-nucleon inelastic
cross section $\sigma_{NN}^{inel}$ are set as 41.6, 67.6 and 68.4 mb for
$\sqrt{s_{NN}}$=0.2, 5.02 and 5.44 TeV\cite{loiz}, respectively. The results
are given in the Tabs. 1-5. In these table, the corresponding improved
MC-Glauber model results \cite{loiz} are also given for the comparisons. We
see in these tables that, the optical Glauber model results are well consistent with corresponding MC-Glauber ones within the error bar, except the most
peripheral collisions.

\begin{center}
\begin{figure}[htbp]
\centering
\hspace{-0.50cm}
\includegraphics[width=0.7\textwidth]{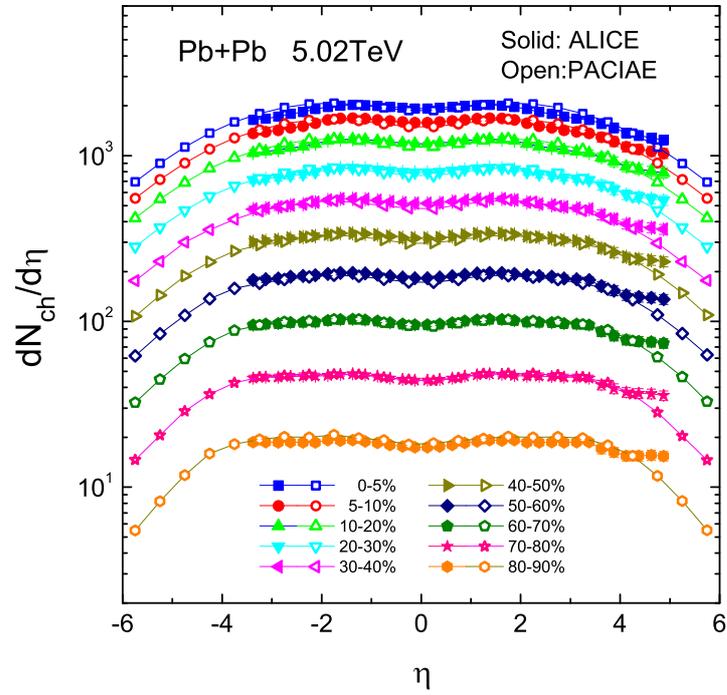}
\caption{(Color online) PACIAE model simulated charged-particle pseudorapidity
distribution in Pb-Pb collisions at $\sqrt{s_{NN}}$=5.02 TeV (solid symbols)
compares with the corresponding ALICE data (open symbols) \cite{ALICE2} (open
symbols).}
\label{eta}
\end{figure}
\end{center}

The PACIAE model simulated charged-particle pseudorapidity distributions (open
symbols) in Pb-Pb collision at $\sqrt{s_{NN}}$=5.02 TeV are compared with the
corresponding ALICE data \cite{ALICE2} (solid symbols) in Fig. \ref{eta} for
ten centrality bins of 0-5\%, 5-10\%, 10-20\%, ..., 80-90\%. This figure shows
that the PACIAE 2.2.2 model results well reproduce the ALICE data from central
to peripheral Pb-Pb collisions. In the $|\eta|>4$ region, the theoretical
results are smaller than the experimental data. This has to by studied further
in the next work by adjusting the parameters in Lund fragmentation function
and/or trying other fragmentation functions.

\begin{center}
\begin{figure}[htbp]
\centering
\hspace{-0.50cm}
\includegraphics[width=0.7\textwidth]{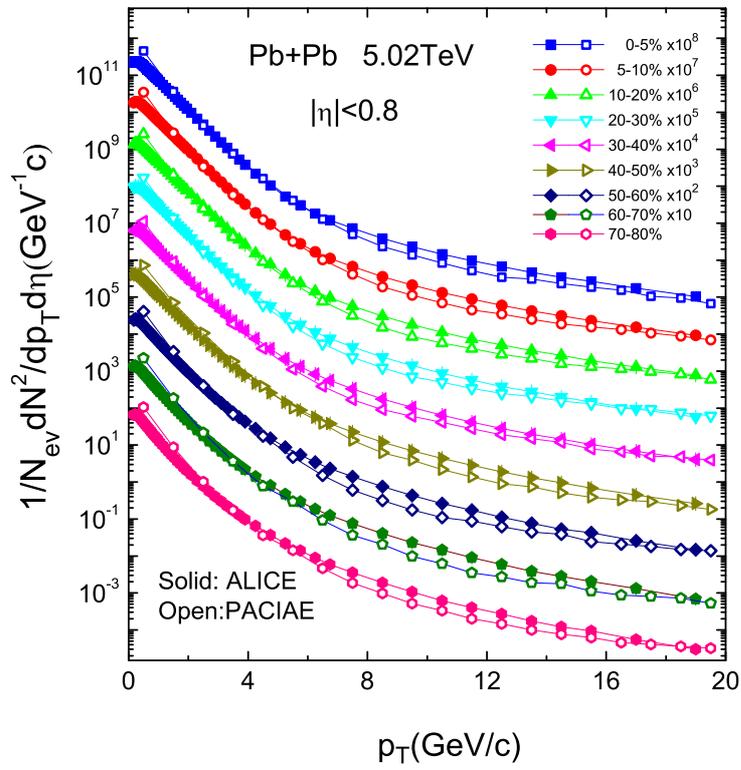}
\caption{(Color online) PACIAE model simulated charged-particle transverse
momentum distribution (open symbols) in Pb-Pb collisions at
$\sqrt{s_{NN}}$=5.02 TeV compares with the corresponding ALICE data (solid
symbols)\cite{ALICE3}. For better visibility, the data except 70-80\% are
rescaled by different factors as indicated in legend.}
\label{pt}
\end{figure}
\end{center}

We compare the PACIAE model simulated charged-particle transverse momentum
distribution in Pb-Pb collision at $\sqrt{s_{NN}}$=5.02 TeV to the
corresponding ALICE data \cite{ALICE3} in Fig~\ref{pt} for centrality bins of
0-5\%, 5-10\%, ..., 70-80\%. For better visibility, both the PACIAE results
and the ALICE data, except 70-80\% one, are rescaled by a factor of 10, $10^2$,
..., $10^8$, respectively. In this figure we see the PACIAE model well
reproduce the corresponding ALICE data in $\pT<$6 GeV/c region. However in the
$\pT>$6 GeV/c region, the theoretical results are smaller than the
experimental data. This might because the $\pT$ distribution in
heavy-ion collisions change from exponential-like at low $\pT$ region
to a power-law shape at high $\pT$ region \cite{ALICE4}, and the $\pT$ distribution
is sampled by exponential-like distribution in this work. In the next study we
shall try to sample particle $\pT$ by power law distribution instead of
exponential-like one in the $\pT>$ 6 GeV/c region.

\section{Summary}
The $f$ coefficient in the impact parameter $b_{max}$ formula in
PACIAE 2.2.2 model is reset to respond the ALICE, ATLAS, and CMS's observation
of the maximum impact parameter in heavy-ion collisions at relativistic
energies should be extended to 20 $fm$. Consequently, the PACIAE model
is updated to a new version of PACIAE 2.2.2 with convenience of studying the
elementary nuclear collisions, proton-nucleus collisions, and the
nucleus-nucleus collisions in an unified program version. The PACIAE model
calculated impact parameter bin and the optical Glauber results of $\Npart$
and $\Ncoll$ are well consistent with the corresponding improved MC-Glauber
ones in the Pb-Pb, $p$-Pb, Xe-Xe, Au-Au, and Cu-Cu collisions at relativistic
energies. PACIAE model simulated charged-particle pseudorapidity and
transverse momentum distributions in Pb-Pb collisions at $\sqrt{s_{NN}}$=5.02
TeV well reproduce the corresponding ALICE data. These results prove the
correction of the impact parameter centrality definition in the updated PACIAE
model and the efficiency of PACIAE model itself.

\textbf{Acknowledgments}
This work was supported by the National Natural Science
Foundation of China under grant Nos.: 11775094, and 11905188 and by the
111 project of the foreign expert bureau of China. YLY acknowledge the financial support from
the Continuous Basic Scientific Research Project (No, WDJC-2019-13) in CIAE.
\label{}




\begin{table}
\centering
\begin{varwidth}{\textwidth}
\caption{Impact parameter $b$, $\langle N_{\rm{coll}}\rangle$ and $\langle N_{\rm{part}}\rangle$ of optical
 Glauber (PACIAE) model for Pb-Pb at $\sqrt{s_{NN}}$=5.02 TeV, and compared with the ones of Monte Carlo Glauber model\cite{loiz}.}
\end{varwidth}

\begin{tabularx}{\textwidth}{CCCCC|CCCC}
\hline
\hline
   & \multicolumn{4}{c|}{optical Glauber (PACIAE) model} & \multicolumn{4}{c} {Monte Carlo Glauber (MCG) model$^{*}$} \\
\hline
Centrality & $b_{\rm{min}}(\rm{fm})$ & $b_{\rm{max}}(\rm{fm})$ & $\langle N_{\rm{coll}}\rangle$ & $\langle N_{\rm{part}}\rangle $
& $b_{\rm{min}}(\rm{fm})$ & $b_{\rm{max}}(\rm{fm})$ & $\langle N_{\rm{coll}}\rangle \pm$rms & $\langle N_{\rm{part}}\rangle \pm$rms \\
\hline
0-5\%    &   0    &  3.46  & 1871.8  & 375.4  & 0 & 3.49 & 1762$\pm$147 & 384.3$\pm$16.6 \\
5-10\%   &  3.46  &  4.89  & 1447.8  & 317.5  & 3.49 & 4.93 & 1380$\pm$113 & 331.2$\pm$17.7 \\
10-15\%  &  4.89  &  5.99  & 1128.1  & 267.4  & 4.93  & 6.04   &  1088$\pm$93.4     & 283$\pm$16.8    \\
15-20\%  &  5.99  &  6.91  & 876.8   & 224.5  & 6.04 &  6.98   &  855.3$\pm$80.8  &  240.9$\pm$16    \\
20-25\%  &  6.91  &  7.73  & 675.4   & 187.4  & 6.98& 7.8& 667.6$\pm$71.6 &204$\pm$15.3  \\
25-30\%  &  7.73  &  8.47  & 512.9   & 155.1  &  7.8 & 8.55 & 515.7$\pm$63.9 & 171.6$\pm$14.7   \\
30-35\%  &  8.47  &  9.14  & 383.9   & 127.3  &  8.55 & 9.23 & 392.9$\pm$57 & 143.2$\pm$14.1   \\
35-40\%  &  9.14  &  9.78  & 281.3   & 103.0  &  9.23 & 9.87 & 294.5$\pm$50 & 118.3$\pm$13.6     \\
40-45\%  &  9.78  & 10.37  & 201.0   &  82.1  &  9.87 & 10.5 & 216.4$\pm$43.3 & 96.49$\pm$13    \\
45-50\%  & 10.37  & 10.93  & 140.2   &  64.4  &  10.5 & 11 & 155.5$\pm$36.6 & 77.48$\pm$12.4     \\
50-55\%  & 10.93  & 11.46  &  95.0   &  49.4  &  11 & 11.6 & 109.2$\pm$30.2 & 61.19$\pm$11.7    \\
55-60\%  & 11.46  & 11.97  &  62.3   &  36.9  &  11.6 & 12.1 & 74.73$\pm$24.3 & 47.31$\pm$10.9   \\

60-65\%  & 11.97  & 12.46  &  39.5   &  26.8  &  12.1 & 12.6 & 49.88$\pm$19.1 & 35.74$\pm$9.96   \\
65-70\%  & 12.46  & 12.93  &  24.3   &  18.8  &  12.6 & 13.1 & 32.38$\pm$14.7 & 26.26$\pm$8.95   \\
70-75\%  & 12.93  & 13.39  &  14.5   &  12.7  &  13.1 & 13.5 & 20.54$\pm$11.1 & 18.75$\pm$7.79   \\
75-80\%  & 13.39  & 13.82  &   8.5   &   8.3  &  13.5 & 14 & 12.85$\pm$8.16 & 13.09$\pm$6.55  \\
80-85\%  & 13.82  & 14.25  &   4.9   &   5.3  &  14 & 14.4 & 8.006$\pm$5.82 & 9.038$\pm$5.22   \\
85-90\%  & 14.25  & 14.66  &   2.8   &   3.2  &  14.4 & 14.9 & 5.084$\pm$4.08 & 6.304$\pm$3.98   \\
90-95\%  & 14.66  & 15.06  &   1.5   &   1.9  &   14.9 & 15.6 & 3.27$\pm$2.77 & 4.452$\pm$2.86  \\
95-100\% & 15.06  & 15.46  &   0.8   &   1.1  &   15.6 & 20 & 2.035$\pm$1.72 & 3.103$\pm$1.8 \\
\hline
\hline \\
\multicolumn{4}{l}{* Data are taken from \cite{loiz}} \\
\end{tabularx}
\label{tab:PbPb502}
\end{table}

\begin{table}
\centering
\begin{varwidth}{\textwidth}
\caption{Impact parameter $b$, $\langle N_{\rm{coll}}\rangle$ and $\langle N_{\rm{part}}\rangle$ of optical
 Glauber (PACIAE) model for p-Pb at $\sqrt{s_{NN}}$=5.02 TeV, and compared with the ones of Monte Carlo Glauber model\cite{loiz}.}
\end{varwidth}

\begin{tabularx}{\textwidth}{CCCCC|CCCC}
\hline
\hline
   & \multicolumn{4}{c|}{optical Glauber (PACIAE) model} & \multicolumn{4}{c} {Monte Carlo Glauber (MCG) model} \\
\hline
Centrality & $b_{\rm{min}}(\rm{fm})$ & $b_{\rm{max}}(\rm{fm})$ & $\langle N_{\rm{coll}}\rangle$ & $\langle N_{\rm{part}}\rangle $
& $b_{\rm{min}}(\rm{fm})$ & $b_{\rm{max}}(\rm{fm})$ & $\langle N_{\rm{coll}}\rangle \pm$rms & $\langle N_{\rm{part}}\rangle \pm$rms \\
\hline
0-5\%    & 0    & 1.79  & 13.1  & 11.2  &   0 & 1.82 & 13.68$\pm$3.51 & 14.68$\pm$3.51  \\
5-10\%   & 1.79 & 2.53  & 12.5  & 10.7  &   1.82 & 2.58 & 13.11$\pm$3.4 & 14.11$\pm$3.4   \\
10-15\%  & 2.53 & 3.10  & 11.8  & 10.1  &   2.58 & 3.16 & 12.5$\pm$3.3 & 13.5$\pm$3.3   \\
15-20\%  & 3.10 & 3.58  & 11.1  &  9.5  &   3.16 & 3.65 & 11.83$\pm$3.18 & 12.83$\pm$3.18   \\
20-25\%  & 3.58 & 4.00  & 10.3  &  8.9  &   3.65 & 4.08 & 11.13$\pm$3.07 & 12.13$\pm$3.07   \\
25-30\%  & 4.00 & 4.38  &  9.5  &  8.3  &  4.08 & 4.47 & 10.36$\pm$2.96 & 11.36$\pm$2.96    \\
30-35\%  & 4.38 & 4.74  &  8.7  &  7.7  &  4.47 & 4.83 & 9.529$\pm$2.83 & 10.53$\pm$2.83    \\
35-40\%  & 4.74 & 5.06  &  7.9  &  7.0  &  4.83 & 5.16 & 8.646$\pm$2.7 & 9.646$\pm$2.7    \\
40-45\%  & 5.06 & 5.37  &  7.0  &  6.3  &  5.16 & 5.47 & 7.721$\pm$2.57 & 8.721$\pm$2.57    \\
45-50\%  & 5.37 & 5.66  &  6.2  &  5.7  &  5.47 & 5.77 & 6.766$\pm$2.41 & 7.766$\pm$2.41    \\
50-55\%  & 5.66 & 5.94  &  5.4  &  5.0  &  5.77 & 6.05 & 5.836$\pm$2.25 & 6.836$\pm$2.25    \\
55-60\%  & 5.94 & 6.20  &  4.6  &  4.4  &  6.05 & 6.32 & 4.949$\pm$2.07 & 5.949$\pm$2.07    \\
60-65\%  & 6.20 & 6.45  &  3.9  &  3.8  &  6.32 & 6.58 & 4.132$\pm$1.87 & 5.132$\pm$1.87    \\
65-70\%  & 6.45 & 6.70  &  3.3  &  3.3  &  6.58 & 6.84 & 3.415$\pm$1.66 & 4.415$\pm$1.66    \\
70-75\%  & 6.70 & 6.93  &  2.7  &  2.8  &  6.84 & 7.1 & 2.802$\pm$1.45 & 3.802$\pm$1.45    \\
75-80\%  & 6.93 & 7.16  &  2.2  &  2.4  &  7.1 & 7.36 & 2.294$\pm$1.23 & 3.294$\pm$1.23    \\
80-85\%  & 7.16 & 7.38  &  1.8  &  2.0  &  7.36 & 7.65 & 1.877$\pm$1.00 & 2.877$\pm$1.00    \\
85-90\%  & 7.38 & 7.59  &  1.5  &  1.7  &  7.65 & 7.99 & 1.55$\pm$0.78 & 2.55$\pm$0.78     \\
90-95\%  & 7.59 & 7.80  &  1.2  &  1.4  &  7.99 & 8.49 & 1.287$\pm$0.56 & 2.287$\pm$0.56    \\
95-100\% & 7.80 & 8.00  &  0.9  &  1.1  &  8.49 & 14.7 & 1.082$\pm$0.30 & 2.082$\pm$0.30    \\
\hline
\hline \\
\end{tabularx}
\label{tab:pPb502}
\end{table}

\begin{table}
\centering
\begin{varwidth}{\textwidth}
\caption{Impact parameter $b$, $\langle N_{\rm{coll}}\rangle$ and $\langle N_{\rm{part}}\rangle$ of optical
 Glauber (PACIAE) model for Xe-Xe at $\sqrt{s_{NN}}$=5.44 TeV, and compared with the ones of Monte Carlo Glauber model\cite{loiz}.}
\end{varwidth}

\begin{tabularx}{\textwidth}{CCCCC|CCCC}
\hline
\hline
   & \multicolumn{4}{c|}{optical Glauber (PACIAE) model} & \multicolumn{4}{c} {Monte Carlo Glauber (MCG) model} \\
\hline
Centrality & $b_{\rm{min}}(\rm{fm})$ & $b_{\rm{max}}(\rm{fm})$ & $\langle N_{\rm{coll}}\rangle$ & $\langle N_{\rm{part}}\rangle $
& $b_{\rm{min}}(\rm{fm})$ & $b_{\rm{max}}(\rm{fm})$ & $\langle N_{\rm{coll}}\rangle \pm$rms & $\langle N_{\rm{part}}\rangle \pm$rms \\
\hline
0-5\%    &  0    &  3.03  & 992.5  & 234.4  &  0 & 3.01 & 942.5$\pm$92.1 & 236.5$\pm$10    \\
5-10\%   &  3.03 &  4.29  & 764.3  & 198.4  & 3.01 & 4.26 & 734.1$\pm$72.8 & 206.1$\pm$11.7    \\
10-15\%  &  4.29 &  5.25  & 591.6  & 166.9  &  4.26 & 5.22 & 571.9$\pm$62 & 177.1$\pm$12.2   \\
15-20\%  &  5.25 &  6.06  & 457.2  & 140.0  &  5.22 & 6.02 & 443.9$\pm$55.5 & 151.1$\pm$12.4   \\
20-25\%  &  6.06 &  6.78  & 348.5  & 116.3  &  6.02 & 6.73 & 341.7$\pm$50.8 & 127.9$\pm$12.6   \\
25-30\%  &  6.78 &  7.43  & 262.3  &  95.8  &  6.73 & 7.38 & 260.5$\pm$46.2 & 107.4$\pm$12.6   \\
30-35\%  &  7.43 &  8.02  & 194.5  &  78.2  &  7.38 & 7.97 & 196.1$\pm$41.7 & 89.36$\pm$12.6   \\
35-40\%  &  8.02 &  8.58  & 141.4  &  63.0  &  7.97 & 8.52 & 145.5$\pm$36.8 & 73.53$\pm$12.4   \\
40-45\%  &  8.58 &  9.10  & 100.4  &  49.9  &  8.52 & 9.04 & 106.5$\pm$31.7 & 59.75$\pm$12.1   \\
45-50\%  &  9.10 &  9.59  &  69.8  &  38.9  &  9.04 & 9.53 & 76.83$\pm$26.8 & 47.94$\pm$11.6   \\
50-55\%  &  9.59 & 10.06  &  47.3  &  29.7  &  9.53 & 9.99 & 54.64$\pm$22.1 & 37.9$\pm$10.9   \\
55-60\%  & 10.06 & 10.50  &  31.4  &  22.2  &  9.99 & 10.4 & 38.28$\pm$18 & 29.43$\pm$10.1   \\
60-65\%  & 10.50 & 10.93  &  20.4  &  16.1  &  10.4 & 10.9 & 26.61$\pm$14.4 & 22.56$\pm$9.17   \\
65-70\%  & 10.93 & 11.35  &  12.8  &  11.4  &  10.9 & 11.3 & 18.25$\pm$11.3 & 16.98$\pm$8.06   \\
70-75\%  & 11.35 & 11.74  &   8.0  &   7.9  &  11.3 & 11.7 & 12.49$\pm$8.7 & 12.68$\pm$6.89   \\
75-80\%  & 11.74 & 12.13  &   4.9  &   5.3  &  11.7 & 12.1 & 8.627$\pm$6.62 & 9.503$\pm$5.74   \\
80-85\%  & 12.13 & 12.50  &   3.0  &   3.4  &  12.1 & 12.5 & 6.011$\pm$4.93 & 7.152$\pm$4.61   \\
85-90\%  & 12.50 & 12.86  &   1.8  &   2.2  &  12.5 & 13.1 & 4.232$\pm$3.64 & 5.422$\pm$3.6   \\
90-95\%  & 12.86 & 13.22  &   1.1  &   1.4  &  13.1 & 13.8 & 2.967$\pm$2.58 & 4.116$\pm$2.67   \\
95-100\% & 13.22 & 13.56  &   0.6  &   0.9  &  13.8 & 20 & 1.95$\pm$1.64 & 3.007$\pm$1.72   \\

\hline
\hline \\
\end{tabularx}
\label{tab:XeXe}
\end{table}

\begin{table}
\centering
\begin{varwidth}{\textwidth}
\caption{Impact parameter $b$, $\langle N_{\rm{coll}}\rangle$ and $\langle N_{\rm{part}}\rangle$ of optical
 Glauber (PACIAE) model for Au-Au at $\sqrt{s_{NN}}$=0.2 TeV, and compared with the ones of Monte Carlo Glauber model\cite{loiz}.}
\end{varwidth}

\begin{tabularx}{\textwidth}{CCCCC|CCCC}
\hline
\hline
   & \multicolumn{4}{c|}{optical Glauber (PACIAE) model} & \multicolumn{4}{c} {Monte Carlo Glauber (MCG) model} \\
\hline
Centrality & $b_{\rm{min}}(\rm{fm})$ & $b_{\rm{max}}(\rm{fm})$ & $\langle N_{\rm{coll}}\rangle$ & $\langle N_{\rm{part}}\rangle $
& $b_{\rm{min}}(\rm{fm})$ & $b_{\rm{max}}(\rm{fm})$ & $\langle N_{\rm{coll}}\rangle \pm$rms & $\langle N_{\rm{part}}\rangle \pm$rms \\
\hline
0-5\%    &   0     &  3.31  & 1075.5  & 345.5  &  0 & 3.31 & 1053$\pm$92.2 & 351$\pm$17.8    \\
5-10\%   &  3.31   &  4.68  &  842.8  & 289.6  &  3.31 & 4.68 & 831.4$\pm$72.1 & 298.1$\pm$17    \\
10-15\%  &  4.68   &  5.74  &  664.6  & 243.0  &  4.68 & 5.73 & 660.1$\pm$61 & 252.7$\pm$16    \\
15-20\%  &  5.74   &  6.62  &  523.3  & 203.8  &  5.73 & 6.61 & 523$\pm$54.4 & 213.8$\pm$15.4    \\
20-25\%  &  6.62   &  7.40  &  409.7  & 170.2  &  6.61 & 7.39 & 412$\pm$49.5 & 180.1$\pm$14.9    \\
25-30\%  &  7.40   &  8.11  &  316.8  & 141.0  &  7.39 & 8.1 & 321.1$\pm$45.3 & 150.8$\pm$14.6    \\
30-35\%  &  8.11   &  8.76  &  241.5  & 115.7  &  8.1 & 8.75 & 247.2$\pm$41.3 & 125.1$\pm$14.3    \\
35-40\%  &  8.76   &  9.37  &  180.9  &  93.8  &  8.75 & 9.35 & 187.8$\pm$37 & 102.8$\pm$13.9    \\
40-45\%  &  9.37   &  9.93  &  133.1  &  75.1  &  9.35 & 9.92 & 139.9$\pm$32.5 & 83.36$\pm$13.4    \\
45-50\%  &  9.93   & 10.47  &   95.8  &  59.2  &  9.92 & 10.5 & 102.4$\pm$27.8 & 66.65$\pm$12.7    \\
50-55\%  & 10.47   & 10.98  &   67.1  &  45.6  &  10.5 & 11 & 73.35$\pm$23.4 & 52.37$\pm$11.9    \\
55-60\%  & 10.98   & 11.47  &   45.8  &  34.4  &  11 & 11.5 & 51.45$\pm$19.2 & 40.39$\pm$11    \\
60-65\%  & 11.47   & 11.94  &   30.3  &  25.2  &  11.5 & 11.9 & 35.33$\pm$15.4 & 30.5$\pm$9.95    \\
65-70\%  & 11.94   & 12.39  &   19.6  &  18.0  &  11.9 & 12.4 & 23.74$\pm$12 & 22.5$\pm$8.79    \\
70-75\%  & 12.39   & 12.82  &   12.3  &  12.4  &  12.4 & 12.8 & 15.64$\pm$9.17 & 16.23$\pm$7.5    \\
75-80\%  & 12.82   & 13.24  &    7.6  &   8.4  &  12.8 & 13.2 & 10.22$\pm$6.83 & 11.55$\pm$6.17    \\
80-85\%  & 13.24   & 13.65  &    4.6  &   5.4  &  13.2 & 13.7 & 6.699$\pm$4.96 & 8.193$\pm$4.86    \\
85-90\%  & 13.65   & 14.05  &    2.7  &   3.5  &  13.7 & 14.2 & 4.426$\pm$3.49 & 5.852$\pm$3.67    \\
90-95\%  & 14.05   & 14.43  &    1.6  &   2.1  &  14.2 & 14.9 & 2.949$\pm$2.38 & 4.216$\pm$2.6    \\
95-100\% & 14.43   & 14.81  &    0.9  &   1.3  &  14.9 & 20 & 1.867$\pm$1.43 & 2.957$\pm$1.57    \\
\hline
\hline \\
\end{tabularx}
\label{tab:AuAu}
\end{table}

\begin{table}
\centering
\begin{varwidth}{\textwidth}
\caption{Impact parameter $b$, $\langle N_{\rm{coll}}\rangle$ and $\langle N_{\rm{part}}\rangle$ of optical
 Glauber (PACIAE) model for Cu-Cu at $\sqrt{s_{NN}}$=0.2 TeV, and compared with the ones of Monte Carlo Glauber model\cite{loiz}.}
\end{varwidth}

\begin{tabularx}{\textwidth}{CCCCC|CCCC}
\hline
\hline
   & \multicolumn{4}{c|}{optical Glauber (PACIAE) model$^{*}$} & \multicolumn{4}{c} {Monte Carlo Glauber (MCG) model} \\
\hline
Centrality & $b_{\rm{min}}(\rm{fm})$ & $b_{\rm{max}}(\rm{fm})$ & $\langle N_{\rm{coll}}\rangle$ & $\langle N_{\rm{part}}\rangle $
& $b_{\rm{min}}(\rm{fm})$ & $b_{\rm{max}}(\rm{fm})$ & $\langle N_{\rm{coll}}\rangle \pm$rms & $\langle N_{\rm{part}}\rangle \pm$rms \\
\hline
0-5\%    &   0     &  2.44  & 217.9  & 105.3  & 0 & 2.34 & 203.6$\pm$24.9 & 106.5$\pm$6.21   \\
5-10\%   &  2.44   &  3.45  & 166.7  &  86.8  & 2.34 & 3.31 & 162.9$\pm$20.6 & 91.68$\pm$6.41   \\
10-15\%  &  3.45   &  4.23  & 127.5  &  71.3  & 3.31 & 4.06 & 130.1$\pm$18 & 78.42$\pm$6.52   \\
15-20\%  &  4.23   &  4.88  &  97.0  &  58.3  & 4.06 & 4.68 & 103.7$\pm$16.3 & 66.83$\pm$6.65   \\
20-25\%  &  4.88   &  5.46  &  73.2  &  47.3  & 4.68 & 5.24 & 82.13$\pm$15 & 56.58$\pm$6.78   \\
25-30\%  &  5.46   &  5.98  &  54.5  &  38.0  & 5.24 & 5.73 & 64.7$\pm$13.8 & 47.63$\pm$6.86   \\
30-35\%  &  5.98   &  6.46  &  40.0  &  30.2  & 5.73 & 6.19 & 50.63$\pm$12.5 & 39.83$\pm$6.86   \\
35-40\%  &  6.46   &  6.90  &  28.9  &  23.6  & 6.19 & 6.62 & 39.28$\pm$11.3 & 33.03$\pm$6.8   \\
40-45\%  &  6.90   &  7.32  &  20.5  &  18.1  & 6.62 & 7.02 & 30.23$\pm$10.2 & 27.14$\pm$6.66   \\
45-50\%  &  7.32   &  7.72  &  14.3  &  13.7  & 7.02 & 7.4 & 23.11$\pm$8.95 & 22.11$\pm$6.43   \\
50-55\%  &  7.72   &  8.09  &   9.9  &  10.2  & 7.4 & 7.77 & 17.54$\pm$7.79 & 17.84$\pm$6.08   \\
55-60\%  &  8.09   &  8.45  &   6.7  &   7.5  & 7.77 & 8.11 & 13.25$\pm$6.69 & 14.3$\pm$5.65   \\
60-65\%  &  8.45   &  8.80  &   4.5  &   5.3  &  8.11 & 8.45 & 9.988$\pm$5.67 & 11.4$\pm$5.13  \\
65-70\%  &  8.80   &  9.13  &   3.0  &   3.8  &  8.45 & 8.78 & 7.576$\pm$4.75 & 9.111$\pm$4.56  \\
70-75\%  &  9.13   &  9.45  &   2.0  &   2.6  &  8.78 & 9.11 & 5.774$\pm$3.9 & 7.305$\pm$3.94  \\
75-80\%  &  9.45   &  9.76  &   1.3  &   1.8  &  9.11 & 9.47 & 4.453$\pm$3.18 & 5.906$\pm$3.34  \\
80-85\%  &  9.76   & 10.06  &   0.8  &   1.2  &  9.47 & 9.86 & 3.465$\pm$2.55 & 4.822$\pm$2.78  \\
85-90\%  & 10.06   & 10.35  &   0.5  &   0.8  &  9.86 & 10.3 & 2.703$\pm$2 & 3.953$\pm$2.23  \\
90-95\%  & 10.35   & 10.64  &   0.3  &   0.5  &  10.3 & 11 & 2.116$\pm$1.52 & 3.261$\pm$1.7  \\
95-100\% & 10.64   & 10.91  &   0.2  &   0.4  &  11 & 19.1 & 1.582$\pm$1.06 & 2.629$\pm$1.15  \\
\hline
\hline \\
\multicolumn{8}{l}{* The tail of the nuclear density profile $d=0.488 fm$ for the Cu.} \\
\end{tabularx}
\label{tab:CuCu}
\end{table}




\begin{thebibliography}{99}

\bibitem{rhic1}
I. Arsene, et al., BRAHMS Collaboration, Nucl. Phys. A 757 (2005) 1.
\bibitem{rhic2}
B. B. Back, et al.,PHOBOS Collaboration, Nucl. Phys. A 757 (2005) 28.
\bibitem{rhic3}
J. Admas, et al., STAR Collaboration, Nucl. Phys. A 757 (2005) 102.
\bibitem{rhic4}
K. Adcox, et al., PHENIX Collaboration, Nucl. Phys. A 757 (2005) 184.
\bibitem{ALICE1}
K. Aamodt, et al., ALICE Collaboration, Phys. Lett. B 696 (2011) 30.
\bibitem{CMS1}
S. Chatrchyan, et al., CMS Collaboration, Eur. Phys. J. C 72 (2012) 1945.
\bibitem{ATLAS1}
G. Aad, et al., ATLAS Collaboration, Phys. Rev. C 86 (2012) 014907.
\bibitem{abel1}
B. I. Abelev, et al., ALICE  Collaboration, Phys. Rev. C 88 (2013) 044909.
\bibitem{sa}
Ben-Hao Sa, Dai-Mei Zhou, Yu-Liang Yan, Xiao-Mei Li, Sheng-Qin Feng,
Bao-Guo Dong, and Xu Cai, Comput. Phys. Comm. 183 (2012) 333.
\bibitem{sa2}
Ben-Hao Sa, Dai-Mei Zhou, Yu-Liang Yan, Bao-Guo Dong, and Xu Cai, Comput. Phys. Comm. 184 (2013) 1476.
\bibitem{zhou} Dai-Mei Zhou, Yu-Liang Yan, Xing-Long Li, Xiao-Mei Li, Bao-Guo Dong, Xu Cai, and Ben-Hao Sa, Comput. Phys. Comm. 193 (2015) 89.
\bibitem{PYTHIA}
Sj\"ostrand T, Mrenna S, Skands P, J. High Energy Phys., 05 (2006) 026.
\bibitem{Combridge}
B.L. Combridge, J. Kripfgang, J. Ranft, Phys. Lett. B 70 (1977) 234.
\bibitem{Field}
R.D. Field, Applications of Perturbative QCD, Addison-Wesley Publishing Company,
Inc., Reading, 1989.
\bibitem{star}
B. I. Abelev, et al.,STAR Collab., Phys. Rev. C 79 (2009) 034909.	
\bibitem{loiz}
C. Loizides, J. Kamin, and D. d'Enterria, Phys. Rev. C 97 (2018) 054910.
\bibitem{esk}
K.J. Eskola, K. Kajantie and J. Lindfors, Nucl. Phys. B 323 (1989) 37;
D. Miskowiec, http://www-linux.gsi.de/~misko/overlap.
\bibitem{Shalid} Amos de Shalid, Herman Feshbach, Nuclear Structure, Theoretical Nuclear
Physics, vol. I, John Wiley and Sons, Inc., New York, 1974.
\bibitem{ALICE2}
J. Adam, et al., ALICE Collaboration, Phys.Lett. B 772 (2017) 567 .
\bibitem{ALICE3}
S. Acharya, et al., ALICE Collaboration, J. High Energy Phys., 1811 (2018) 013
\bibitem{ALICE4}
Adam J. et al., ALICE Collaboration, Phys. Rev. C 93 (2016) 034913S.

\end{thebibliography}
\end{document}